\documentclass[11pt]{article}
\usepackage{amsmath,amssymb,bm,epsf,epsfig,graphicx}
\input epsf.sty
\usepackage[utf8]{inputenc}
\usepackage[english]{babel}
\usepackage{amsmath}
\usepackage{latexsym}
\usepackage{amsfonts}
\usepackage{mathtools}
\usepackage{amssymb}
\usepackage{scalerel}
\usepackage{extpfeil}
\usepackage[left=3cm,right=3cm,top=3.1cm,bottom=3.1cm]{geometry}
\usepackage{cite}
\usepackage{tensor}
\usepackage{tikz}
\usepackage{tikz-cd}
\usepackage[mathscr]{euscript}
\usetikzlibrary{arrows.meta, bending}
\tikzset{C/.style={circle, minimum size=8mm,
		node contents={},
		append after command={\pgfextra{%
				\draw[-{Straight Barb[flex']}](\tikzlastnode.150) arc (450:110:2.8mm);}
	}}
}

\usepackage{hyperref}
\numberwithin{equation}{section}

\def\aver#1{\left\langle\, #1 \,\right\rangle}

\def\p{\partial}

\def \be {\begin{eqnarray}}
\def \ee {\end{eqnarray}}
\def \bal {\begin{align}}
\def \eal {\end{align}}
\def \bdm {\begin{displaymath}}
\def \edm {\end{displaymath}}

\def\0{\nonumber}

\begin{document}
	\begingroup\allowdisplaybreaks

\vspace*{1.1cm}

\centerline{\Large \bf Closed string deformations in open string field theory III: }\vspace{.3cm}
 \centerline{\Large \bf ${\cal N}=2$ worldsheet localization}
\vspace*{.1cm}

\begin{center}

{\large Carlo Maccaferri$^{(a)}$\footnote{Email: maccafer at gmail.com} and Jakub Vo\v{s}mera$^{(b)}$\footnote{Email: jvosmera at phys.ethz.ch} }
\vskip 1 cm
$^{(a)}${\it Dipartimento di Fisica, Universit\`a di Torino, \\INFN  Sezione di Torino and Arnold-Regge Center\\
Via Pietro Giuria 1, I-10125 Torino, Italy}
\vskip .5 cm
$^{(b)}${\it Institut f\"{u}r Theoretische Physik, ETH Z\"{u}rich\\
	Wolfgang-Pauli-Straße 27, 8093 Z\"{u}rich, Switzerland}

\end{center}

\vspace*{6.0ex}

\centerline{\bf Abstract}
\bigskip
In this paper, which is the last of a series including \cite{OC-I, OC-II} we first verify that the two  open-closed effective potentials  derived in the previous paper from the WZW theory in the large Hilbert space and the $A_\infty$ theory in the small Hilbert space have the  same vacuum structure. In particular, we show that mass-term deformations given by the effective (open)$^2$-closed couplings are the same, provided the effective tadpole is vanishing to first order in the closed string deformation. We show that this condition is always realized when the worldsheet BCFT enjoys a global ${\cal N}=2$ superconformal symmetry and the deforming closed string belongs to the chiral ring in both the holomorphic and anti-holomorphic sector. In this case it is possible to explicitly evaluate the mass deformation by localizing the SFT Feynman diagrams to the boundary of world-sheet moduli space, reducing the amplitude to a simple open string two-point function.  As a non-trivial check of our construction we couple a constant Kalb-Ramond closed string state to the OSFT on the $\text{D}3$--$\text{D}(-1)$ system and we show that half of the bosonic blowing-up moduli  become tachyonic, making the system  condense to a bound state whose binding energy we compute exactly to second order in the closed string deformation, finding agreement with the literature.

\baselineskip=16pt
\newpage
\tableofcontents

\section{Introduction and summary}

This paper is devoted to the evaluation of the  open-closed effective couplings derived in \cite{OC-II} in the WZW theory in the large Hilbert space \cite{Berkovits, Berkovits:2004xh}  and in the related  $A_\infty$ theory in the small Hilbert space \cite{Erler:2013xta}.
There, deforming the pure open string field theory with an elementary open/closed coupling and integrating out the open string fields outside the kernel of $L_0$ (the ``massive'' open string fields if we work at zero momentum) we have  obtained  effective couplings between  arbitrary number of on-shell closed string states (representing the deformation) and  dynamical open strings in the kernel of $L_0$. Formally these effective couplings are physical string theory amplitudes  computed respectively in the large Hilbert space (for the WZW theory) and in the small Hilbert space (for the $A_\infty$ theory). Naively, it should not matter in which theory we compute these amplitudes  because if two theories are related by field redefinition (as it is the case for the partially gauge-fixed WZW theory and the $A_\infty$ theory) they should give rise to the same amplitudes. However this statement, like many others based on the ``canceled propagator'' argument (i.e. the fact that the propagator inverts $Q_{\rm BRST}$), is only valid for S-matrix elements computed at generic values of the external momenta. Zero momentum amplitudes, like the ones which give rise to algebraic couplings in the effective action, are an important exception because they receive non-trivial contributions from the boundary of moduli space. Indeed it turns out that the open-closed amplitudes of the WZW and $A_\infty$ theory  in general differ from contributions at the boundary of moduli space, which are in general not vanishing at zero momentum. At a given order,  a sufficient condition for the equality of two amplitudes with the same external states is that all lower order amplitudes vanish\footnote{A consequence of this fact is that exactly marginal deformations coincide in the two theories.}.  In fact this was already (implicitly) observed in \cite{Maccaferri:2019ogq} where the same problem (without the closed string deformation) was discussed. There it was shown that the quartic effective massless couplings obtained from the WZW and the $A_\infty$ theory were the same, provided the cubic couplings were vanishing. This may seem surprising but in fact  it is perfectly consistent with the possibility of doing field redefinitions at zero momentum while staying in the massless cohomology, and it essentially means that zero momentum amplitudes should be really thought of as pieces of an effective potential which is obviously defined up to (constant) field redefinitions. In this paper we are mainly interested in (open)$^2$-(closed) amplitudes giving rise to mass-term deformations\footnote{Analogous couplings for open strings in the R sector have been obtained in \cite{Billo:2008sp}.}
, so the only relevant lower order amplitude is just the effective (open)-(closed) tadpole coupling a single closed string with a single massless open string. But such a massless tadpole has to vanish anyway in order to allow for a stable vacuum in perturbation theory and so the mass terms are guaranteed to coincide in the two theories. 

The possibility of dealing with amplitudes at  zero (open string) momentum is an advantage of the (open) string field theory framework, which gives a unique consistent way to deal with divergences coming from  open string degeneration using the open string propagator \cite{Larocca:2017pbo, Sen:2019jpm, Sen:2020cef}, something which is not possible in usual string perturbation theory. 
In the case of the superstring, this consistent treatment of the zero momentum sector allowed to discover surprising properties of some amplitudes. Indeed, inspired by a  computation in heterotic string theory (section 8 of \cite{Sen-restoration}), it was shown in \cite{Erbin:2019spp,  Maccaferri:2018vwo} that the effective quartic coupling of purely open or closed massless strings can often be computed by localization  at the boundary of the world-sheet moduli space  reducing the four-point amplitude  to a simple two-point function (on the disk for open strings and on the sphere for closed strings). The key property which ensures this mechanism is the global ${\cal N}=2$ superconformal symmetry (the same which gives rise to space-time supersymmetry) and the fact that the states which enter into the amplitude belong to the chiral ring, i.e.\ they sit in the short multiplets being annihilated by one of the two supercurrents and are charged under the ${\cal N}=2$ R-charge. This mechanism directly applies in our open-closed setting as well, provided the deforming closed string state belongs to the chiral ring  in both holomorphic sectors (and thus is a NS-NS state) and combines with the quartic coupling of the massless open string already computed in \cite{Maccaferri:2018vwo}. When some of the massless fields become tachyonic due to the closed string deformation, the obtained potential becomes Mexican-hat-like and it identifies a new less energetic vacuum for the initial D-brane system which, to the given order in the closed string deformation, can be analytically characterised. 

The paper is organized as follows. In section \ref{subsec:AinfBerk} we discuss the relation between the open-closed coupling of the WZW and $A_\infty$ theory and in particular we show that when the leading contribution to the effective tadpole is vanishing, then the leading order contribution to the mass deformation is the same in both theories. In section \ref{sec:6} we switch to the ${\cal N}=2$ setup where we first show that if the deformed closed string belongs to the chiral ring, its effective tadpole necessarily vanishes to first order. Subsequently we show that the (open)$^2$-(closed) coupling simplifies and it reduces to a boundary contribution in moduli space given by a sum of two-point functions of auxiliary fields associated to the open and closed strings.  In section \ref{sec:7} we put to test our construction by switching on a Kalb-Ramond deformation in the $\text{D3}$-$\overline{\text{D}(-1)}$ system, which perfectly fit in the ${\cal N}=2$ setting and therefore has a vanishing massless tadpole. Appendix \ref{app:proofs} contains some detailed calculations which we use in the main text.

\section{WZW-like vs $A_\infty$ effective potentials
}
\label{subsec:AinfBerk}

Let us now focus on comparing the effective couplings for states $\psi\in\text{ker}\,L_0\equiv P_0\mathcal{H}$ obtained from the WZW-like SFT and $A_\infty$ microscopic SFTs modified by adding the Ellwood invariant in the previous paper \cite{OC-II} of the series. We will do so explicitly only for a couple of lowest orders in perturbation theory, making no attempt to construct a full field redefinition relating the WZW-like and $A_\infty$ SFT effective actions. Taking a lesson from the field redefinition \cite{Erler:2015rra} relating these two theories at the microscopic level, it is to be expected that the leading order $\mathcal{O}(\psi^1, \mu^0)$ of the field redefinition for the effective actions will be trivial. Below we will verify this expectation by showing (for a couple of lowest orders) that the effective potentials predicted by the two SFTs consistently agree at any given order provided that  lower-order effective couplings vanish. In other words, we check that at zero-momentum, the two theories yield effective potentials which impose identical constraints on exactly marginal directions.

\subsection{WZW-like and $A_\infty$ effective actions}

Recall that in the previous paper \cite{OC-II} of the series, we have found that the effective $A_\infty$ action in marginal closed-string background, up to a non-dynamical cosmological constant, takes the form
\begin{align}
\tilde{S}^{(\mu)}(\lambda\psi) = \sum_{k=0}^\infty\sum_{\alpha=0}^\infty \lambda^{k+1}\mu^\alpha\tilde{S}_{k,\alpha}(\psi)\,,
\end{align}
where the 
$k+1$ open-string and $\alpha$ closed-string coupling $\tilde{S}_{k,\alpha}(\psi)$ can be expressed as
\begin{align}
\tilde{S}_{k,\alpha}(\psi)=\frac{1}{k+1}\omega_\text{S}(\psi,\tilde{N}_{k,\alpha}(\psi^{\otimes k}))\,,\label{eq:Ska}
\end{align}
with cyclic products $\tilde{N}_{k,\alpha}$ that can be extracted from an all-order expression given in the coalgebra language. For the first couple of lowest orders, we explicitly obtain
\begin{subequations}
	\begin{align}
	\tilde{N}_{0,1}&=P_0e\,,\\
	\tilde{N}_{0,2}&=-P_0E_1(he)+P_0M_2(he,he)\,,\\
	\tilde{N}_{1,0}(\psi)&=P_0Q\psi\,,\\
	\tilde{N}_{1,1}(\psi)&=P_0E_1(\psi)-P_0M_2(he,\psi)-P_0M_2(\psi,he)\,,\\
	\tilde{N}_{2,0}(\psi,\psi)&=P_0M_2(\psi,\psi)\,,\\
	\tilde{N}_{3,0}(\psi,\psi)&=P_0M_3(\psi,\psi,\psi)-P_0M_2(\psi,hM_2(\psi,\psi))-P_0M_2(hM_2(\psi,\psi),\psi)\,,\\
	&\hspace{0.2cm}\vdots\nonumber
	\end{align}
\end{subequations}
where we have taken $h=(b_0/L_0)\bar{P}_0$ to be the propagator for the modes outside of $\text{ker}\,L_0$ in the Siegel gauge and $P_0$ to be the projector onto $\text{ker}\,L_0$.
At the same time, the analogous WZW-like effective action was found to read (again, up to a non-dynamical cosmological constant)
\be
S_{\rm eff}^{(\mu)}(\lambda\varphi)&=&
\lambda \left(\mu\aver{e,\varphi}-\frac{\mu^2}{2}\aver{[he,\tilde h e],\varphi}+O\left(\mu^3\right)\right)\0\\
&+&\lambda^2\left(\frac12\aver{\eta\varphi,Q\varphi}-\frac{\mu}{2}\aver{\eta\varphi,\left[\tilde h he,Q\varphi\right]}+O\left(\mu^2\right)\right)\0\\
&+&\lambda^3\left(-\frac16\aver{\eta\varphi,[\varphi,Q\varphi]}+O\left(\mu\right)\right)\0\\
&+&\lambda^4\left(\frac1{24}\aver{\eta\varphi,[\varphi,[\varphi,Q\varphi]]}-\frac18\aver{[\eta\varphi,Q\varphi],\tilde h h[\eta\varphi,Q\varphi]}+O(\mu)\right)\0\\
&+&O(\lambda^5)\,,\label{eq:SeffWZW}
\ee
where $\tilde{h} = [\xi_0 - X_0(b_0/L_0)]\bar{P}_0$ denotes the dual propagator introduced in \cite{OC-II}, which fixes the $\eta$-gauge symmetry for the massive modes by setting $\xi_0\bar{P}_0\Phi=0$ provided that the Siegel gauge condition $b_0\bar{P}_0\Phi=0$ is simultaneously imposed. Note that in order to facilitate comparison between the two theories, we also fix the $\eta$-gauge symmetry of the effective WZW-like action by assuming $\xi_0\varphi=0$, that is $\varphi=\xi_0\psi$. 

\subsection{Open SFT effective vertices}

Starting with the purely open effective couplings, the kinetic term
\begin{align}
\tilde{S}_{1,0}(\psi) = \frac{1}{2}\omega_\text{S}(\psi,Q\psi) = \frac{1}{2}\big\langle \psi,Q\psi\big\rangle_\text{S}
\end{align}
is clearly identical in both theories. For the cubic effective coupling, the $A_\infty$ theory yields
\begin{align}
\tilde{S}_{2,0}(\psi) = \frac{1}{3}\omega_\text{S}(X_0\psi,m_2(\psi,\psi)) = \frac{1}{3}\big\langle X_0 \psi,\psi^2\big\rangle_\text{S}
\end{align}
while the WZW-like theory gives the cubic coupling
\begin{align}
\frac{1}{3}\big\langle X_0\psi,\psi^2\big\rangle_\text{S}
-\frac{1}{6}\big\langle \psi,[\psi,\xi_0Q\psi]+[\xi_0\psi,Q\psi]\big\rangle_\text{S}\,.
\end{align}
Hence, we observe that the effective cubic couplings in the two theories coincide if $\psi\in \text{ker}\,L_0$ is physical, that is, if
\begin{align}
P_0 Q\psi\equiv \tilde{N}_{1,0}(\psi)=0
\end{align}
and we have $\tilde{S}_{1,0}(\psi)=0$. On the other hand, in agreement with the results of \cite{Erler:2015rra,Erler:2015uba,Erler:2015uoa}, it is easy to check that in the general case when $Q\psi\neq 0$, the WZW-like cubic coupling is obtained from the $A_\infty$ cubic coupling by performing the corresponding field redefinition. Moreover, if in addition $\psi$ satisfies the projector condition
\begin{align}
P_0 M_2(\psi,\psi)\equiv\tilde{N}_{2,0}(\psi,\psi)=0 \,,\label{eq:PC}
\end{align}
we have $\tilde{S}_{2,0}(\psi)=0$. As discussed in \cite{Vosmera:2019mzw}, such a condition can be realized e.g.\ by assuming $\psi$ to be of the form 
\begin{align}
\psi = c\mathbb{V}_\frac{1}{2} e^{-\phi}\,,
\end{align}
where $\mathbb{V}_{1/2}$ is a $h=1/2$ zero-momentum matter state belonging to the (anti)-chiral ring of a worldsheet $\mathcal{N}=2$ superconformal algebra. It was then shown in \cite{Maccaferri:2019ogq} that upon assuming the projector condition \eqref{eq:PC}, the quartic couplings $\tilde{S}_{3,0}(\psi)$ computed using the WZW-like and $A_\infty$ effective actions agree. Hence, in line with our expectation that the leading order of the field redefinition relating the effective WZW-like and $A_\infty$ theories is trivial,
we confirm (at least to the quartic order) that the purely open effective couplings in the two theories at a fixed order in perturbation theory agree, provided that lower order couplings vanish. Since the vanishing of the effective potential order-by-order in perturbation theory is precisely the requirement on massless modes to be exact moduli for any given background, we conclude that both the WZW-like and the $A_\infty$ theory yield the same algebraic constraints determining moduli spaces of open string backgrounds. 

\subsection{Open-closed effective vertices}

Let us now extend this line of thought to include open-closed couplings generated by deforming the microscopic action with Ellwood invariant. Starting with the tadpole coupling $\tilde{S}_{0,1}(\psi)$, this is clearly the same in the two theories. Moreover, assuming that the projector condition 
\begin{align}
P_0 e\equiv \tilde{N}_{0,1}=0\label{eq:PCe}
\end{align}
holds, we clearly have $\tilde{S}_{0,1}(\psi)=0$. This is saying that at the leading order in $\mu$, the effective tadpole vanishes, so that the vacuum shift can be written as 
$\psi_\mu=\mathcal{O}(\mu^2)$ and therefore we obtain a consistent leading-order microscopic vacuum shift 
\begin{align}
\Psi_\mu=\mu h e +\mathcal{O}(\mu^2)\,.
\end{align}
Similarly to the case of the open-string projector condition \eqref{eq:PC}, we will see in Subsection \ref{subsec:tadpole} below that the projector condition \eqref{eq:PCe} is guaranteed to hold if the bosonic Ellwood state takes the form 
\begin{align}
e=\varepsilon_{ij}\big[X_0U^i(i){U}^j(-i)\big] I+
\varepsilon_{ij}\big[U^i(i){X}_0{U}^j(-i)\big] I\,,\label{eq:e0}
\end{align}
where we have introduced chiral Fock states
\begin{align}
U^i &=c\mathbb{U}^i_\frac{1}{2} e^{-\phi}\,,
\end{align}
with $\mathbb{U}_{1/2}^i$ some zero-momentum matter states with $h=1/2$ and  belonging to the chiral ring of some global worldsheet $\mathcal{N}=2$ superconformal algebra. As usual, $I$ denotes the identity string field. Also, $\varepsilon_{ij}$ is some polarization and we will assume that it already includes the action of the gluing automorphism (fixed by the open-string background) which arises when we map the original closed string field to a bi-local field the upper half-plane.

Let us now proceed with showing that, assuming the projector conditions $\tilde{N}_{0,1}=P_0 e=0$ and $\tilde{N}_{1,0}(\psi)=P_0 Q\psi=0$, the coupling $\tilde{S}_{1,1}(\psi)$ computed in the $A_\infty$ SFT agrees with the one computed in the WZW-like theory. This coupling of course determines the leading order correction to the mass-term due to a non-zero closed string vev at zero momentum.
Note that restricting to the on-shell fields $\psi$ is physically well justified, as it can be shown that the algebraic correction to the Siegel gauge propagator $h=(b_0/L_0)\bar{P}_0$ coming from integrating out the Nakanishi-Lautrup field does not contribute at this order in perturbation theory -- see \cite{Erbin:2020eyc} for an analogous result in the bosonic string. Starting on the $A_\infty$ side, we first extract an explicit expression for the 1-product $\tilde{N}_{1,1}$, namely
\begin{align}
\tilde{N}_{1,1}(\psi)=P_0 E_1(\psi)-P_0M_2(he,\psi)-P_0M_2(\psi,h e)\,.
\end{align}
We will now continue by rewriting $\tilde{S}_{1,1}(\psi)$ in terms of the bosonic products and the PCO zero mode $X_0$ assuming that $Q\psi=0=P_0e$. Starting with the expression
	\begin{align}
	\tilde{S}_{1,1}(\psi) 
		&= \frac{1}{2}\omega_\text{S}(\psi,E_1(\psi))-\frac{1}{2}\omega_\text{S}(\psi,M_2(h e,\psi))-\frac{1}{2}\omega_\text{S}(\psi,M_2(\psi,h e))\,.\label{eq:S11}
\end{align}
for the $\mathcal{O}(\mu)$ mass-term correction, it is then shown in detail in Appendix \ref{app:proofS11} that expanding the superstring products $E_1$ and $M_2$ in the large Hilbert space in terms of the gauge products and eventually the bosonic products with explicit PCO insertions, the expression \eqref{eq:S11} can be rewritten as
\begin{align}
\tilde{S}_{1,1}(\psi)=\frac{1}{2}\omega_{\text{S}}( m_2(X_0\psi,\psi),h e)+\frac{1}{2}\omega_{\text{S}}( m_2(\psi,X_0\psi),h e)\,.\label{eq:MassTermCorrA}
\end{align}
Finally, in order to facilitate the comparison with the WZW-like theory, we can rewrite \eqref{eq:MassTermCorrA} using the unsuspended notation in the large Hilbert space, namely\footnote{Recall that we define 
\begin{align}
	\varphi = \xi_0 c \mathbb{V}_\frac{1}{2} e^{-\phi}\,,
\end{align}
as well as
	\begin{align}
	\mathcal{U}^i &=\xi_0c\mathbb{U}_\frac{1}{2}^ie^{-\phi}\,.
	\end{align}
}
\begin{align}
	\tilde{S}_{1,1}(\eta\varphi) &=-\frac{1}{2}\big\langle [\eta\varphi,Q\varphi],\tilde{h}h e\big\rangle\,,\label{eq:S11Berk}
\end{align}
which indeed agrees with the result \eqref{eq:SeffWZW} we obtained in the previous paper \cite{OC-II} from the WZW-like theory. 

Similarly, we can show that the subleading contribution to the effective tadpole $\tilde{S}_{0,2}(\psi)$ computed in the $A_\infty$ theory agrees with the corresponding result obtained in the WZW-like theory provided that the conditions $\tilde{N}_{1,0}(\psi)=P_0Q\psi=0$ and $\tilde{N}_{0,1}=P_0e=0$ are satisfied. Here we start with the 0-product
\begin{align}
\tilde{N}_{0,2} = -P_0 E_1(h e)+P_0 M_2(h e, h e)\,,
\end{align}
which yields the subleading effective tadpole
\begin{align}
\tilde{S}_{0,2}(\psi) 
&= -\omega_\text{S}(\psi,E_1(h e))+\omega_\text{S}(\psi,M_2(h e,h e))\,.
\end{align}
We then verify in detail in Appendix \ref{app:proofS02} that after expanding the superstring products $E_1$ and $M_2$ in terms of the bosonic products, the coupling $\tilde{S}_{0,2}(\psi)$ can be rewritten as
\begin{align}
\tilde{S}_{0,2}(\psi)&= +\frac{1}{2}\big\langle\xi_0 \psi, [
Q\xi_0 he,he]\big\rangle\,.\label{eq:tS02}
\end{align}
Finally, we can rewrite this in terms of the partially gauge-fixed string field $\varphi=\xi_0\psi$ and the dual propagator $\tilde{h}$ as
\begin{align}
\tilde{S}_{0,2}	&= -\frac{1}{2}\big\langle\varphi, [
\tilde{h} e,h e]\big\rangle\,,
\end{align}
so that again we obtain exact agreement with the result \eqref{eq:SeffWZW} of the WZW-like SFT.

These findings clearly constitute evidence that the two effective SFT describe the same perturbative physics for any given background. In order to characterize the field redefinition relating the two effective SFTs to all orders, one will first have to get a firm handle on the structure of the effective WZW-like vertices in a future project.
 
\section{${\cal N}=2$ localization of open-closed couplings}\label{sec:6}
In this section we will argue that the computation of certain effective open-closed couplings can be simplified considerably provided the background admits a certain $\mathcal{N}=2$ superconformal symmetry. We will thereby generalize the results of \cite{Maccaferri:2018vwo} (and subsequently of \cite{Maccaferri:2019ogq,Vosmera:2019mzw,Erbin:2019spp}) by including on-shell closed string insertions. In particular, we will see that in the cases where the background supports a global worldsheet $\mathcal{N}=2$ superconformal symmetry such that all matter states with $h=1/2$ carry $U(1)$ $R$-charge $q=\pm 1$, the leading-order tadpole $\tilde{S}_{0,1}$ vanishes and the leading-order mass-term correction $\tilde{S}_{1,1}$ localizes on the boundary of the bosonic worldsheet moduli space, where the propagator degenerates to an infinitely long strip. 

\subsection{$R$-charge decomposition of matter fields}

In more detail, we will assume that we can pick a global $\mathcal{N}=2$ superconformal algebra $\{T,G^{\pm},J\}$ with the above-described features in relation to its spectrum of irreducible representations. We can then decompose any $h=1/2$ matter state as
\begin{align}
	\mathbb{V}_\frac{1}{2} &= \mathbb{V}_\frac{1}{2}^++ \mathbb{V}_\frac{1}{2}^-\,,	
\end{align}
and correspondingly (writing $\varphi = \xi_0 c \mathbb{V}_\frac{1}{2} e^{-\phi}$)
\begin{align}
	\varphi = \varphi^+ + \varphi^-\,,
\end{align}
where the states with superscript $\pm$ carry $U(1)$ $R$-charge $\pm 1$.
We will further assume that both left- and right-moving components of the on-shell closed string insertion can be decomposed as
	\begin{align}
		\mathbb{U}^i_\frac{1}{2} &=(\mathbb{U}^i_\frac{1}{2})^+ +(\mathbb{U}^i_\frac{1}{2})^-\,,
	\end{align}
and correspondingly, calling
	$\mathcal{U}^i =\xi_0c\mathbb{U}_\frac{1}{2}^ie^{-\phi}\,$

	\begin{align}
		\mathcal{U}^i &=(\mathcal{U}^i)^+ +(\mathcal{U}^i)^-\,.
	\end{align}
We then have
\begin{align}
	J_0 \mathbb{V}_\frac{1}{2}^\pm &= \pm \mathbb{V}_\frac{1}{2}^\pm\,,
\end{align}
together with
\begin{subequations}
	\begin{align}
		G^{\pm}(z)\mathbb{V}_\frac{1}{2}^\mp(0) &=\frac{1}{z}\mathbb{V}_1^\mp(0)+\text{reg.}\,,\\
		G^{\pm}(z)\mathbb{V}^\pm_\frac{1}{2}(0) &= \text{reg.}\,,
	\end{align}
\end{subequations}
where $J_0 \mathbb{V}_1^\pm=0$ and we have denoted $|\mathbb{V}_1^\mp\rangle = G^\pm_{-\frac{1}{2}}|\mathbb{V}^\mp_\frac{1}{2}\rangle$. Also note that we have
\begin{align}
	Q\varphi^\pm &= c\mathbb{V}_1^\pm-\gamma \mathbb{V}_\frac{1}{2}^\pm\,,
\end{align}
where $\gamma = e^{\phi}\eta$. Similarly for $\mathbb{U}_\frac{1}{2}^i$.

\subsection{Computing the leading-order tadpole}
\label{subsec:tadpole}

Let us first consider the leading-order effective tadpole coupling $\tilde{S}_{0,1}(\psi)=\omega_\text{S}(\psi,e)$. Expanding $e$ in terms of the on-shell closed string state in picture $-1$, we can write
\begin{align}
\tilde{S}_{0,1}(\psi) = \varepsilon_{ij}\big\langle\psi,[(X_0 U^i)(i){U}^j(-i)]I\big\rangle+\varepsilon_{ij}\big\langle\psi,[ U^i(i)({X}_0{U}^j)(-i)]I\big\rangle\,.
\end{align}
Using the form $\psi=c\mathbb{V}_{1/2}e^{-\phi}$, the matter part of the coupling $\tilde{S}_{0,1}(\psi)$ is therefore proportional to the boundary 3-point functions
\begin{subequations}
\begin{align}
&\big\langle \mathbb{V}_\frac{1}{2}(0)\,[G_{-\frac{1}{2}}\mathbb{U}^i_\frac{1}{2}](i)\,{\mathbb{U}}^k_\frac{1}{2}(-i)\big\rangle\,,\\
&\big\langle \mathbb{V}_\frac{1}{2}(0)\,\mathbb{U}^i_\frac{1}{2}(i)\,[G_{-\frac{1}{2}}{\mathbb{U}}^k_1](-i)\big\rangle\,,
\end{align}
\end{subequations}
where $G$ denotes the $\mathcal{N}=1$ (local) worldsheet supercurrent.
However, it was argued in \cite{Vosmera:2019mzw,Erbin:2019spp} that such correlators always vanish whenever the background supports a global worldsheet $\mathcal{N}=2$ superconformal symmetry generated by $\{T,G^\pm,J\}$ such that $G=G^++G^-$ and all matter states with $h=1/2$ carry $U(1)$ $R$-charge $q=\pm 1$. Hence, we conclude that in such cases the leading-order tadpole $\tilde{S}_{0,1}(\psi)$ vanishes. Moreover, we can prove a somewhat stronger result, namely that $\tilde{N}_{0,1}=P_0 e$ vanishes. Using that $|I\rangle =|0\rangle+L_{-2}|0\rangle+\ldots$, we eventually obtain
\begin{align}
P_0 e = -\varepsilon_{ij}\,c\p c\,\Big(\big\{[G_{-\frac{1}{2}}\mathbb{U}_\frac{1}{2}^i] {\mathbb{U}}_{\frac{1}{2}}^j\big\}_1+\big\{\mathbb{U}_{\frac{1}{2}}^i  [G_{-\frac{1}{2}}{\mathbb{U}}_\frac{1}{2}^j]\big\}_1\Big)\, e^{-\phi}\,,
\end{align}
where $\{AB\}_k$ denotes the field in the symmetric OPE of $A$ with $B$ at the order of singularity $z^{-k}$. However, it was again argued in \cite{Vosmera:2019mzw,Erbin:2019spp} that the OPE of $G_{-\frac{1}{2}}\mathbb{V}_\frac{1}{2}$ with ${\mathbb{W}}_{\frac{1}{2}}$ does not have integer poles whenever the local $\mathcal{N}=1$ superconformal symmetry generated by $\{T,G\}$ enhances to a global worldsheet $\mathcal{N}=2$ superconformal symmetry with the above-described parameters. In such cases, we therefore have $P_0 e=0$.

\subsection{Localizing the leading-order mass-term correction}

Let us now consider the leading-order mass-term correction $\tilde{S}_{1,1}$ which arises after turning on the $\mu$-deformation. Starting with the expression \eqref{eq:S11Berk} for $\tilde{S}_{1,1}$ in the large Hilbert space, we can first rewrite this as
\begin{align}
\tilde{S}_{1,1}(\eta\varphi) &=\frac{1}{2}\varepsilon_{ij}\bigg\langle[\eta \varphi, Q\varphi ],\xi_0
\frac{b_0}{L_0}\bar{P}_0[(\eta\mathcal{U}^i)(i)({Q}{\mathcal{U}}^j)(-i)]I\bigg\rangle+\nonumber\\
&\hspace{4cm}+\frac{1}{2}\varepsilon_{ij}\bigg\langle[ \eta \varphi,Q\varphi ]
,\xi_0\frac{b_0}{L_0}\bar{P}_0[(Q\mathcal{U}^i)(i)({\eta}{\mathcal{U}}^j)(-i)]I\bigg\rangle\,.\label{eq:amp}
\end{align}
This is analogous to the Berkovits-like form of the quartic effective action of \cite{Maccaferri:2019ogq} and the Berkovits-like form of the obstruction at third order of \cite{Vosmera:2019mzw}.
The amplitude \eqref{eq:amp} contains one Siegel-gauge propagator insertion (and therefore has to be evaluated by integrating over one bosonic worldsheet modulus -- the length of the Siegel-gauge strip), it is shown in detail in Appendix \ref{app:proofloc} using the techniques developed by \cite{Maccaferri:2018vwo} that $\tilde{S}_{1,1}$ can be recast as (suppressing from now on the insertion points of the closed string fields at $\pm i$ for the sake of clarity)
\begin{align}
	\tilde{{S}}_{1,1}(\eta\varphi)
	&=-\frac{1}{2}\varepsilon_{ij}\big\langle\big([ \eta \varphi^+,Q\varphi^- ]\!-\![ \eta \varphi^-,Q\varphi^+ ]\big)
	,\!{P}_0\big[(\mathcal{U}^i)^+ ({\mathcal{U}}^j)^- \!-\!(\mathcal{U}^i)^- ({\mathcal{U}}^j)^+\big]I\big\rangle\nonumber\\
	&\hspace{2.5cm}-\frac{1}{2}\varepsilon_{ij}\big\langle[  \varphi^+,Q\varphi^+ ]
	,{P}_0\big[(\mathcal{U}^i)^-{\eta}({\mathcal{U}}^j)^- - {\eta}(\mathcal{U}^i)^-({\mathcal{U}}^j)^-\big]I\big\rangle\nonumber\\
	&\hspace{2.5cm}-\frac{1}{2}\varepsilon_{ij}\big\langle[  \varphi^-,Q\varphi^- ]
	,{P}_0\big[(\mathcal{U}^i)^+{\eta}({\mathcal{U}}^j)^+ - {\eta}(\mathcal{U}^i)^+({\mathcal{U}}^j)^+\big]I\big\rangle\,.\label{eq:S11loc}
\end{align}
That is, the computation of $\tilde{S}_{1,1}$ localizes on the boundary of the bosonic worldsheet moduli space, where the Siegel-gauge strip degenerates to an infinite length.
Let us now define the closed-string auxiliary fields $\mathbb{G}_1^\pm$, $\mathbb{G}_0$ through the bulk-boundary OPE (which of course depends on the superconformal boundary condition we are imposing)
\begin{subequations}
	\label{eq:defAux1}
	\begin{align}
		\mathbb{G}_1^\pm &=\varepsilon_{ij}\lim_{z\to {z}^\ast}\left[(\mathbb{U}_\frac{1}{2}^i)^\pm (z)({\mathbb{U}}_\frac{1}{2}^j)^\pm ({z}^\ast)\right]\,,\\
		\mathbb{G}_0 &=\varepsilon_{ij}\lim_{z\to {z}^\ast}\left[2z\left(
		(\mathbb{U}_\frac{1}{2}^i)^- (z)({\mathbb{U}}_\frac{1}{2}^j)^+ ({z}^\ast)-(\mathbb{U}_\frac{1}{2}^i)^+ (z)({\mathbb{U}}_\frac{1}{2}^j)^- ({z}^\ast)\right)\right]\,.
	\end{align}
\end{subequations}
We also recall the definitions of the open-string auxiliary fields from \cite{Maccaferri:2018vwo}
\begin{subequations}
	\label{eq:defAux2}
	\begin{align}
		{\mathbb{H}}_1^\pm &= \lim_{z\to 0}\Big[\mathbb{V}_\frac{1}{2}^\pm(z){\mathbb{V}}_\frac{1}{2}^\pm(-z) \Big]\,,\\
		{\mathbb{H}}_0&=\lim_{z\to 0}\Big[2z\Big(\mathbb{V}_{\frac{1}{2}}^-(z) {\mathbb{V}}_{\frac{1}{2}}^+(-z)-{\mathbb{V}}_{\frac{1}{2}}^+(z)\mathbb{V}_{\frac{1}{2}}^-(-z)\Big)\Big]  \,,
	\end{align}
\end{subequations}
It is then straightforward to show that 
\begin{subequations}
	\begin{align}
		P_0\big([ \eta \varphi^+,Q\varphi^- ]\!-\![ \eta \varphi^-,Q\varphi^+ ]\big) &= +2\eta c\, {\mathbb{H}}_0+\ldots\,,\\[+1.2mm]
		P_0[  \varphi^\pm,Q\varphi^\pm ] &= -2c\, {\mathbb{H}}_1^\pm+\ldots\,,\\[+1mm]
		{P}_0\varepsilon_{ij}\big[(\mathcal{U}^i)^\pm {\eta}({\mathcal{U}}^j)^\pm - {\eta}(\mathcal{U}^i)^\pm({\mathcal{U}}^j)^\pm\big]I &=+2\xi c\p c e^{-2\phi}\mathbb{G}_1^\pm +\ldots\,,\\[+1.0mm]
		{P}_0\varepsilon_{ij}\big[(\mathcal{U}^i)^+ ({\mathcal{U}}^j)^- \!-\!(\mathcal{U}^i)^- ({\mathcal{U}}^j)^+\big]I &=-\xi\p \xi c\p c  e^{-2\phi}\mathbb{G}_0+\ldots\,,
	\end{align}
\end{subequations}
where $\ldots$ denote possible additional terms which, however, do not contribute into the final expression for the effective coupling.
We can therefore write
\begin{align}
	\tilde{{S}}_{1,1} = 2\langle\mathbb{G}_1^-| {\mathbb{H}}_1^+\rangle+2\langle\mathbb{G}_1^+| {\mathbb{H}}_1^-\rangle+\langle\mathbb{G}_0| {\mathbb{H}}_0\rangle\,,
	\label{eq:ObstCoh}
\end{align}
where we have noted that
\begin{subequations}
	\begin{align}
		\langle c\p c\xi e^{-2\phi}(z)c(w)\rangle_\mathrm{L} &= -(z-w)^2\,,\\
		\langle c\p c\xi\p \xi e^{-2\phi}(z)c\eta (w)\rangle_\mathrm{L} &= -1\,.
	\end{align}
\end{subequations}
Recalling the expression for the (4-open,0-closed) coupling $\tilde{{S}}_{30}$
\begin{align}
	\tilde{{S}}_{3,0}=\langle\mathbb{H}_1^+| {\mathbb{H}}_1^-\rangle+\frac{1}{4}\langle\mathbb{H}_0| {\mathbb{H}}_0\rangle
\end{align}
reported in \cite{Maccaferri:2018vwo}, it is evident that the (4-open,0-closed) and (2-open,1-closed) terms contribute into the effective action $\tilde{{S}}^{(\mu)}$ at zero momentum as (completing squares)
\begin{subequations}
	\begin{align}
		\tilde{{S}}^{(\mu)}&\supset \tilde{{S}}_{3,0}+ \mu \tilde{{S}}_{1,1}\\
		&=\langle\mathbb{H}_1^++2\mu\mathbb{G}_1^+| {\mathbb{H}}_1^-+2\mu\mathbb{G}_1^-\rangle+\frac{1}{4}\langle\mathbb{H}_0+2\mu\mathbb{G}_0| {\mathbb{H}}_0+2\mu\mathbb{G}_0\rangle+\nonumber\\
		&\hspace{6cm}-4\mu^2\Big(\langle\mathbb{G}_1^+| \mathbb{G}_1^-\rangle+\frac{1}{4}\langle \mathbb{G}_0|\mathbb{G}_0\rangle\Big)\,.\label{eq:above1}
	\end{align}
\end{subequations}
Carefully inspecting the definitions \eqref{eq:defAux1} and \eqref{eq:defAux2} (and assuming we are working in a unitary sector of the total matter CFT), all terms in the expression \eqref{eq:above1} can be seen to be negative semi-definite (where semi- is accounting for the possibility that our set of auxiliary fields may not be linearly independent). This shows that provided that the \emph{(deformed) generalized ADHM constraints} 
\begin{subequations}
	\label{eq:genADHMbulk}
	\begin{align}
		\mathbb{H}_1^\pm &=-2\mu\mathbb{G}_1^\pm+\mathcal{O}(\mu^2)\,,\\	
		\mathbb{H}_0 &=-2\mu\mathbb{G}_0\,+\mathcal{O}(\mu^2)\,,
	\end{align}
\end{subequations}
are solvable, the corresponding matter fields $(\mathbb{V}_{1/2}^\pm)^\ast$ (and the respective string field $\varphi^\ast$) determine, to the given order in perturbation theory, a global minimum of the effective potential 
\begin{align}
\tilde{V}^{(\mu)} = -\tilde{{S}}^{(\mu)}\,.\label{eq:above12}
\end{align}
This clearly occurs in the case when the positive semi-definite undeformed quartic potential $-\tilde{{S}}_{3,0}$ in \eqref{eq:above1} is non-zero and the mass-term correction $\mu\tilde{{S}}_{1,1}$ renders some of the modes tachyonic. Also note that while the conditions \eqref{eq:genADHMbulk} are clearly sufficient for obtaining a global minimum of the effective potential, the semi-definitness of \eqref{eq:above1} means that the generalized ADHM constraints \eqref{eq:genADHMbulk} are not always necessary, namely they can be overconstraining. While this does not happen in the example of the $\text{D}(-1)/\text{D}3$ system, which is analyzed in detail below, it turns out that there exist other backgrounds for which the inner product $\langle \cdot | \cdot\rangle$ can become degenerate on the set of auxiliary fields $\{\mathbb{H}_1^\pm,\mathbb{H}_0\}$.
An example of such a background is provided by the $\text{D}(-1)/\text{D}7$ system recently investigated by \cite{Billo:2021xzh}.

The effective potential \eqref{eq:above12} then has a generalized Mexican-hat shape and gives the \emph{tachyon potential} (to given order in perturbation theory) for such modes. Since the string field $\varphi^\ast$ then solves the effective (massless) equation of motion and we have argued that there are no additional constraints coming from out-of-gauge equations for $R$, it follows that $\Phi^\ast\equiv \Phi^{(\mu)}(\varphi^\ast)=\varphi^\ast+R^{(\mu)}(\varphi^\ast)$ solves the full SFT equation of motion. Provided that they are solvable, the equations \eqref{eq:genADHMbulk} therefore determine a classical solution of the full SFT corresponding to a barely relevant deformation.
Also note that since the perturbative vacuum\footnote{The choice $\varphi=0$ indeed corresponds to the \emph{true} perturbative vacuum of the effective SFT (to the given order in $\mu$) because the tadpole term vanishes at leading order in $\mu$.} $\mathbb{H}_1^\pm=\mathbb{H}_0=0$  clearly gives $\tilde{{S}}^{(\mu)}=0$, the effective potential difference $\Delta\tilde{{V}}^{(\mu)}=-\Delta\tilde{{S}}^{(\mu)}$ between the perturbative vacuum and the classical solution given by \eqref{eq:genADHMbulk} is equal to 
\begin{align}
	\Delta\tilde{{V}}^{(\mu)}
	&= 4\mu^2\bigg(\big\langle \mathbb{G}_1^+\big|\mathbb{G}_1^-\big\rangle+\frac{1}{4}\big\langle\mathbb{G}_0\big|\mathbb{G}_0\big\rangle\bigg)+\mathcal{O}(\mu^3)\equiv\tilde{{V}}^{(\mu)}_\text{min}\,.\label{eq:Vmin}
\end{align}
This computes the difference between the energies of the initial brane configuration and the classical solution (tachyon condensate).

\section{Testing the construction with non-commutative instantons}

Let us now apply these results to a general NSNS deformation of the $\overline{\text{D}(-1)}$/D3 system in flat space. We will denote by $k$  the number of $\overline{\text{D}(-1)}$ branes and by $N$ the number of $\text{D3}$ branes, whose worldvolume we take to be compactified on a 4-torus with unit volume. 
This system has been already studied in a first-quantized setting with the Kalb-Ramond field in the bulk background \cite{Billo:2005fg}.  Moreover  tachyon condensation in a $\text{D}(-1)/\text{D}3$ system with a $B$-field has been also analyzed using the numerical SFT methods (level truncation) in \cite{David:2000um}. So we are in a nice set-up where we can compare our exact SFT approach to existing results. We recall the set-up of \cite{Maccaferri:2018vwo,Mattiello:2019gxc,Vosmera:2019mzw}, where the bosonic $X^1,\ldots,X^4$ and fermionic $\psi^1,\ldots,\psi^4$ target coordinates span the euclidean D3 worldvolume. We will also find it convenient to introduce the complexified fermions
	\begin{align}
		\psi^{r\pm}&=\frac{1}{\sqrt{2}}(\psi^{2r-1}\pm  i \psi^{2r})\,.
	\end{align}
Generally, we will take the latin indices $r,s=1,2$ to run over the complexified coordinates and $\mu,\nu=1,\ldots,4$ over the standard euclidean coordinates. The localizing $U(1)$ $R$-current
can then be expressed as $J=J_1+J_2$
where $J_r =\, :\! \psi_{r-}\psi_{r+}\!:$. We therefore write the matter part of $\mu e$ as
\begin{subequations}
	\label{eq:defpol}
	\begin{align}
		\mu\varepsilon_{\mu\nu}\big[(\mathbb{U}_\frac{1}{2}^\mu)^\pm (i)({\mathbb{U}}_\frac{1}{2}^\nu)^\pm (-i)\big]I &=\mu\varepsilon_{r\pm s\pm}\big[\psi^{r\pm}\, (i){\psi}^{s\pm}(-i)\big]I\,,\\
		\mu\varepsilon_{\mu\nu}\big[(\mathbb{U}_\frac{1}{2}^\mu)^\pm (i)({\mathbb{U}}_\frac{1}{2}^\nu)^\mp (-i)\big]I&=\mu\varepsilon_{r\pm s\mp}\big[\psi^{r\pm}\, (i){\psi}^{s\mp}(-i)\big]I\,,
	\end{align}
\end{subequations}
where $\mathbb{U}_{1/2}^\mu = \psi^\mu$. 
Hence, for $({\mathbb{U}}_{1/2}^\mu)^\pm$ we can write
\begin{subequations}
	\label{eq:trpol}
	\begin{align}
		({\mathbb{U}}_\frac{1}{2}^{2r-1})^\pm  &= +\frac{1}{\sqrt{2}} \psi^{r\pm}\,,\\
		({\mathbb{U}}_\frac{1}{2}^{2r})^\pm  &= \mp\frac{i}{\sqrt{2}}  \psi^{r\pm}\,.
	\end{align}
\end{subequations}
The polarization tensor $\mu\varepsilon_{\mu\nu}$ is determined by the corresponding NSNS field deformation, as well as by the boundary conditions imposed by the $\overline{\text{D}(-1)}$/D3 system.
At leading order in the deformation, we expect to be able to identify (note that we set $\alpha'=1$)
\begin{align}
	\mu\varepsilon_{\mu\nu}= C
	\bigg(
	\begin{array}{cc}
	1_{N\times N} & 0_{N\times k}\\
	0_{k\times N} & -1_{k\times k}
	\end{array}
	\bigg)
	\big(\delta g_{\mu\nu}+2\pi \delta B_{\mu\nu}\big)\label{eq:pol}
\end{align}
where the precise value of the factor $C$ will be determined momentarily. Also note that the gluing-condition dependence of $\varepsilon_{\mu\nu}$ is responsible for the appearance of the Chan-Paton factor in \eqref{eq:pol} where the upper-left Chan-Paton sector is localized on the D3-brane stack and the lower-right corner is localized on the $\overline{\text{D}(-1)}$-brane stack. Let us also isolate the bulk dependence of $\mu\varepsilon_{\mu\nu}$ by stripping the Chan-Paton factor, namely by defining
\begin{align}
	\mu\tilde{\varepsilon}_{\mu\nu}\equiv C
	\big(\delta g_{\mu\nu}+2\pi \delta B_{\mu\nu}\big)\,.\label{eq:mue}
\end{align}
Performing the corresponding bulk-boundary OPE, we when obtain
\begin{subequations}
	\label{eq:GsD1D3}
	\begin{align}
		\mathbb{G}_1^\pm &= \begin{pmatrix}
			\frac{1}{N}1_{N\times N} & 0_{N\times k}\\
			0_{k\times N} & -\frac{1}{k}1_{k\times k}
		\end{pmatrix}:\!\psi^{1\pm}\psi^{2\pm}\!:(\tilde{\varepsilon}_{1\pm 2\pm}-\tilde{\varepsilon}_{2\pm 1\pm})\,,\\
		\mathbb{G}_0 &= \begin{pmatrix}
			\frac{1}{N}1_{N\times N} & 0_{N\times k}\\
			0_{k\times N} & -\frac{1}{k}1_{k\times k}
		\end{pmatrix}\sum_{r=1,2}(\tilde{\varepsilon}_{r- r+}-\tilde{\varepsilon}_{r+ r-})\,.
	\end{align}
\end{subequations}
Note that the relative normalization of the bulk-boundary OPE coefficients in the respective Chan-Paton sectors will prove instrumental for recovering the correct form of the Fayet-Iliopoulos terms in the corresponding 4d $N=2$ gauge theory action (see for instance eq.\ (4.11) of \cite{David:2002wn}). This normalization was fixed so that we obtain correctly normalized 1-point functions of bulk operators on UHP for the stacks of $k$ $\overline{\text{D}(-1)}$ branes and $N$ $\text{D}3$ branes compactified on a 4-torus with unit volume, as one can easily check. It is also immediately clear that the metric deformation does not enter the auxiliary fields $\mathbb{G}_1^\pm$ and $\mathbb{G}_0$ at all (because only the antisymmetric part of the polarization tensor enters \eqref{eq:GsD1D3}): that is, we have $\mathbb{G}_1^\pm = \mathbb{G}_0=0$ when only metric deformations are turned on. Considering the explicit form \eqref{eq:mue} of $\mu\tilde{\varepsilon}_{\mu\nu}$, we can rewrite the bulk auxiliary fields as
\begin{subequations}
	\begin{align}
		2\mu\mathbb{G}_1^\pm &=\hspace{0.8mm}\zeta_\mathbb{C}(\delta B)\hspace{0.8mm}\begin{pmatrix}
			\frac{1}{N}1_{N\times N} & 0_{N\times k}\\
			0_{k\times N} & -\frac{1}{k}1_{k\times k}
		\end{pmatrix}:\!\psi^{1\pm}\psi^{2\pm}\!:\,,\\
		2\mu\mathbb{G}_0 &= 2\zeta_\mathbb{R}(\delta B)\begin{pmatrix}
			\frac{1}{N}1_{N\times N} & 0_{N\times k}\\
			0_{k\times N} & -\frac{1}{k}1_{k\times k}
		\end{pmatrix}\,.
	\end{align}
\end{subequations}
where we have introduced the combinations
\begin{subequations}
	\label{eq:zetas}
	\begin{align}
		\zeta_\mathbb{C}(\delta B)&\equiv 4\pi i C(\delta B_{13}\mp i \delta B_{14}\mp i\delta B_{23}-\delta B_{24})\,,\\
		\zeta_\mathbb{R}(\delta B)&\equiv 4\pi i C(\delta B_{12}+\delta B_{34})\,.
	\end{align}
\end{subequations}
Recalling then the expressions for $\mathbb{H}_1^\pm$, $\mathbb{H}_0$ from \cite{Maccaferri:2018vwo,Vosmera:2019mzw}, which can be computed from the matter field
\begin{align}
	\mathbb{V}_{\frac{1}{2}} &\equiv     \begin{pmatrix}
		A_\mu \psi^\mu & w_{{\alpha}}\Delta S^{{\alpha}}\\
		\bar{w}_{{\alpha}}\bar{\Delta} S^{{\alpha}} & a_\mu \psi^\mu   
	\end{pmatrix}\,,
\end{align}
(where $A_\mu$ is the $U(N)$ gauge field of the D3 branes, $a_\mu$ are the $U(k)$ transverse scalars of the $\overline{\text{D}(-1)}$ branes in the D3 directions and $w_\alpha,\bar{w}_\alpha$ are the $N\times k$ and $k\times N$ stretched string moduli with $\Delta$ and $S^\alpha$ the bosonic and fermionic twist fields -- see \cite{Vosmera:2019mzw} for our conventions), we can finally establish by substituting into the general formula \eqref{eq:above1} that the effective potential can be written as\footnote{Here we also implement the reality conditions $(A_\mu)^\dagger = A_\mu$, $(a_\mu)^\dagger = a_\mu$, $(\bar{w}_\alpha)^\dagger = w^\alpha$ which follow from the reality of the string field; see \cite{Vosmera:2019mzw} for a discussion}
\begin{align}
	\tilde{V}^{(\mu)}-\tilde{V}^{(\mu)}_\mathrm{min} &= +\mathrm{tr}\Big|[A_{1+},A_{2+}]-w_+\bar{w}_++\frac{\zeta_\mathbb{C}(\delta B)}{N}\Big|^2+\nonumber\\[2mm]
	&\hspace{0.4cm}+\mathrm{tr}\Big|[\,a_{1+}\,,\,a_{2+}]+\bar{w}_+ w_+-\frac{\zeta_\mathbb{C}(\delta B)}{N}\Big|^2+\nonumber\\[+2mm]
	&\hspace{0.4cm}+ \frac{1}{4}\mathrm{tr}\Big(\sum_{r=1,2}[(A_{r+})^\dagger,A_{r+}]\!-\! w_+(w_+)^\dagger\! + \! (\bar{w}_+)^\dagger\bar{w}_++\frac{2\zeta_\mathbb{R}(\delta B)}{N}\Big)^2+\nonumber\\
	&\hspace{0.4cm}+ \frac{1}{4}\mathrm{tr}\Big(\sum_{r=1,2}[\,(a_{r+})^\dagger\,,\,a_{r+}] \! -\! \bar{w}_+ (\bar{w}_+)^\dagger\!+\!({w}_+)^\dagger w_+-\frac{2\zeta_\mathbb{R}(\delta B)}{k}\Big)^2+\nonumber\\[1.8mm]
	&\hspace{0.4cm}+\mathcal{O}(\mu^3)\,.\label{eq:N2gauge}
\end{align}
We have therefore recovered the correct F- and D-term structure of the corresponding 4d $N=2$ gauge theory action (see \cite{David:2002wn}) with FI terms $\zeta_\mathbb{C}(\delta B)$ and $\zeta_\mathbb{R}(\delta B)$.
At this point we note that by performing a suitable rotation of the coordinate axes along the D3-brane worldvolume, we can always arrange that $\delta B_{13}=\delta B_{14}=\delta B_{23}=\delta B_{24}=0$ which gives $\zeta_\mathbb{C}(\delta B)=0$. Namely, we can always set $B_{\mu\nu}$ to be proportional to a combination of the selfdual and anti-selfdual t'Hooft symbols $\eta^3_{\mu\nu}$, $\bar{\eta}^3_{\mu\nu}$. Requiring that the effective potential attains a minimum, we then obtain the non-commutative ADHM constraints
\begin{subequations}
	\label{eq:NCADHM}
	\begin{align}
		\mu_\mathbb{C}(a,w,\bar{w})&=\mathcal{O}(\mu^2)\,,\\
		\tilde{\mu}_\mathbb{C}(A,w,\bar{w})&=\mathcal{O}(\mu^2)\,,\\
		\mu_\mathbb{R}(a,w,\bar{w})&=+2\zeta_\mathbb{R}(\delta B)/k+\mathcal{O}(\mu^2)\,,\\ 
		\tilde{\mu}_\mathbb{R}(A,w,\bar{w}) &= -2\zeta_\mathbb{R}(\delta B)/N+\mathcal{O}(\mu^2)\,,
	\end{align}
\end{subequations} 
where we have introduced the real and complex hyper-K\"{a}hler moment maps
\begin{subequations}
	\begin{align}
		\mu^{\mathbb{C}}(a,w,\bar{w}) &\equiv [\,a_{1+}\,,\,a_{2+}]+\bar{w}_+ w_+ \,,\\[+1mm]
		\tilde{\mu}^{\mathbb{C}}(A,w,\bar{w}) &\equiv [A_{1+},A_{2+}]-{w}_+ \bar{w}_+ \,,
	\end{align}
\end{subequations}
as well as 
\begin{subequations}
	\begin{align}
		{\mu}^{\mathbb{R}}(a,w,\bar{w}) &\equiv [\,(a_{r+})^\dagger\,,\,a_{r+}]+(w_+)^\dagger w_+-\bar{w}_+(\bar{w}_+)^\dagger \,,\\[+1mm]
		\tilde{\mu}^{\mathbb{R}}(A,w,\bar{w}) &\equiv [(A_{r+})^\dagger,A_{r+}] -w_+ (w_+)^\dagger + (\bar{w}_+)^\dagger \bar{w}_+ \,.
	\end{align}
\end{subequations}
These results are clearly in agreement with standard literature on the subject \cite{Douglas:1996sw,Seiberg:1999vs,Nekrasov:1998ss,David:2002wn} (where we can recover the more usual sign conventions for ${\mu}^{\mathbb{R}}$, $\tilde{\mu}^{\mathbb{R}}$ by rescaling $A_{r+}\to iA_{r+}$). The constraints \eqref{eq:NCADHM} then determine the moduli space $\mathfrak{M}_{k,N}(\zeta_\mathbb{R})$ of non-commutative instantons whose dimension turns out to be $\dim\mathfrak{M}_{k,N}(\zeta_\mathbb{R})=4kN$. It is also apparent from \eqref{eq:N2gauge} that upon setting $\zeta_\mathbb{C}(\delta B)=0$, only the D-terms contribute to the mass-squared $(m_+)^2$ and $(\bar{m}_+)^2$ of the stretched string modes $w_+$ and $\bar{w}_+$, namely
\begin{subequations}
	\label{eq:masses}
	\begin{align}
		(m_+ )^2 &= -\zeta_{\mathbb{R}}(\delta B)\Big(\frac{1}{k}+\frac{1}{N}\Big)+\mathcal{O}(\mu^2)\,,\\
		(\bar{m}_+)^2 &= +\zeta_{\mathbb{R}}(\delta B)\Big(\frac{1}{k}+\frac{1}{N}\Big)+\mathcal{O}(\mu^2)\,.
	\end{align}
\end{subequations}
It is therefore clear from \eqref{eq:masses} that for $\zeta_\mathbb{R}(\delta B)>0$, we obtain that $w_+$ becomes tachyonic and $\bar{w}_+$ acquires real mass, while for $\zeta_\mathbb{R}(\delta B)<0$ we have that $\bar{w}_+$ becomes tachyonic and ${w}_+$ acquires real mass. By analyzing the stretched string spectrum in the presence of a $B$-field (see e.g.\ \cite{Seiberg:1999vs,David:2000um}) and denoting $2\pi\delta B_{12}\equiv \delta b$, $2\pi\delta B_{34}\equiv \delta b'$, it is not hard to independently find that
\begin{subequations}
	\label{eq:masses2}
	\begin{align}
		(m_+ )^2 &=  -\frac{1}{2\pi}(\delta b+ \delta b')+\mathcal{O}( (\delta B)^2)\,,\\
		(\bar{m}_+ )^2 &=  +\frac{1}{2\pi}(\delta b+\delta b')+\mathcal{O}( (\delta B)^2)\,,
	\end{align}
\end{subequations}
so that comparing \eqref{eq:masses} with \eqref{eq:masses2} (and recalling the definition \eqref{eq:zetas}), it finally follows that we need to identify 
\begin{align}
	C=\frac{1}{4\pi i}\frac{kN}{k+N}\,.
\end{align}
Assuming that the non-commutative ADHM equations \eqref{eq:NCADHM} are solved, we can therefore use \eqref{eq:Vmin} to compute the mass defect
\begin{subequations}
	\begin{align}
		\tilde{V}^{(\mu)}_{\text{min}} &= -(\zeta_\mathbb{R}(\delta B))^2 \Big(\frac{1}{k}+\frac{1}{N}\Big)+\mathcal{O}(\mu^3)\\
		&=-\Big(\frac{\delta b+\delta b'}{2\pi}\Big)^2 \frac{kN}{k+N}+\mathcal{O}((\delta B)^3)\,.\label{eq:SminNC}
	\end{align}
\end{subequations}
The corresponding full SFT solution should therefore be interpreted as a true $\overline{\text{D}(-1)}/\text{D}3$ bound state.
Observe that the result \eqref{eq:SminNC} can be independently obtained from space-time supersymmetry considerations: starting with the expressions\footnote{Our normalization conventions in \eqref{eq:GsD1D3} and below are consistent with taking the D3 branes to be compact with volume set to 1. Also note that we are setting $g_\text{o}=1$.}
\begin{align}
	M_{\text{D}(-1)} &=\frac{1}{2\pi^2}\,,\\
	M_{\text{D}3} &=\frac{1}{2\pi^2}\sqrt{(1+b^2)(1+(b')^2)}\,,
\end{align}
for the masses of a single $\text{D}(-1)$ and D3 brane in a $B$-field, and also noting the expression \cite{Obers:1998fb,David:2000um}
\begin{align}
	M(k,N) = \frac{1}{2\pi^2}\sqrt{[k+(1-bb')N]^2+(b+b')^2 N^2}
\end{align}
for the mass of the bound state of $k$ $\overline{\text{D}(-1)}$ and $N$ $\text{D}3$ branes in a $B$-field (which follows from requiring that the bound state saturates the BPS bound with 8 supercharges \cite{Obers:1998fb}), it is not hard to show that we indeed obtain a mass defect
\begin{align}
	kM_{\text{D}(-1)}+NM_{\text{D}3}-M(k,N) = +\Big(\frac{\delta b+\delta b'}{2\pi}\Big)^2 \frac{kN}{k+N}+\mathcal{O}((\delta B)^3)\,,
\end{align}
which is consistent with the result \eqref{eq:SminNC}.
This independently verifies our worldsheet calculations.

A couple of comments are in order. First, note that provided that we choose the $B$-field to be anti-selfdual (that is $b=-b'$), we obtain
\begin{align}
	\mathbb{G}_0=\mathbb{G}_1^\pm =\zeta_\mathbb{R}(\delta B)= \zeta_\mathbb{C}(\delta B) = m_+ = \bar{m}_+=\tilde{V}^{(\mu)}_{\text{min}}=0\,,
\end{align}
so that the strings stretching between the $\overline{\text{D}(-1)}$ and D3 branes are massless and the system is bound only marginally, as it is well known (see e.g. \cite{Seiberg:1999vs}). Also note that as opposed to the case without a $B$-field, it is a known result \cite{Nekrasov:1998ss,Seiberg:1999vs} that the non-commutative ADHM equations \eqref{eq:NCADHM} have a non-trivial solution even in the case $k=N=1$. For generic values of the $B$-field (where the $\text{D}(-1)/\text{D3}$ system forms a true bound state), it is also clear that the constituent D$p$-branes cannot be separated by an open-string marginal deformation. Generic non-zero $B$-field therefore resolves the small-instanton singularity in the moduli space which normally appears at zero $B$-field. Furthermore, had we considered $\text{D}(-1)$ branes instead of $\overline{\text{D}(-1)}$ branes, the localizing current would instead read $J = J_1 - J_2$ and we would therefore obtain $\zeta_\text{R}(\delta B)=4\pi i C(\delta B_{12}-\delta B_{34})$. The system would then become marginally bound for selfdual values of the $B$-field.

Note that when the $\text{D}(-1)/\text{D3}$ system forms a true bound state, it clearly cannot be described using a boundary state with the conventional Dirichlet or Neumann conditions. This therefore must be an example of an unconventional boundary state which, while being otherwise consistent, does not satisfy any gluing conditions of the type
\begin{subequations}
	\begin{align}
	(\alpha_m^\mu + \tensor{\Omega}{^\mu_\nu}\bar{\alpha}^\nu_{-m})\|\text{D}(-1)/\text{D}3,\eta\rangle\!\rangle &=0\,,\\
	(\psi_r^\mu + i\eta \tensor{\Omega}{^\mu_\nu}\bar{\psi}^\nu_{-r} )\|\text{D}(-1)/\text{D}3,\eta\rangle\!\rangle &=0\,,
	\end{align}
\end{subequations}
where $\tensor{\Omega}{^\mu_\nu}$ is an automorphism of the oscillator algebra.
Given the fact that the bound state saturates the BPS bound while conserving 8 spacetime supercharges, it is to be expected that the corresponding boundary state will satisfy gluing conditions for the (localizing) global $\mathcal{N}=2$ superconformal algebra.
Using the methods developed in this paper, it should be possible to study such boundary states perturbatively around the region in the $B$-field moduli space, where the stretched $\text{D}(-1)/\text{D}3$ strings become massless. It also turns out that it is possible to write down exact analytic expressions for such boundary states at certain points in both bulk and boundary moduli space using a Gepner-like construction \cite{Schnabl:2019oom}.

\section{Discussion and outlook}\label{sec:7}
In this paper, following the construction started in \cite{Erbin:2020eyc, Koyama:2020qfb} and continued in \cite{OC-I,OC-II}, we have first established that the effective potential derived from the WZW theory in the large Hilbert space and the one derived from the $A_\infty$ theory in the small Hilbert space have the same vacuum structure, although they are in general related by a zero momentum field redefinition. Then we have explicitly evaluated the first  superstring open-closed effective couplings using a powerful localization technique which considerably simplifies the evaluation of the involved chiral four-point functions by reducing them to the boundary of the worldsheet moduli space. The method gives a completely universal form for the quartic effective potential which can be used at will for general different backgrounds enjoying a worldsheet ${\cal N}=2$ superconformal symmetry. We expect there is a rich structure that waits to be uncovered in this direction and what we have done up to now  is probably just a small step. Here we offer several possible directions for future research.
\begin{itemize}
\item We have only studied the localization of the mass terms in case of a NS-NS deformation. It would be useful to search for similar mechanisms in case of R-R deformations. 

\item It would be useful, possibly taking advantage of the ${\cal N}=2$ spectral flow, to study what happens to localization after adding the R sector.

\item It would be interesting to study the problems related to closed string degeneration along the lines of what has been proposed in \cite{OC-I} for the bosonic string. In addition to this, when ${\cal N}=2$ superconformal symmetry is present, there is good chance that one could use  analogous localization mechanisms directly in the open-closed superstring field theory \cite{Moosavian:2019ydz}. This would have the advantage of giving us control on both open and closed string degeneration in a full field theoretical way,  while not having to deal with the complicated off-shell structure of the open-closed vertices which  typically deal with the interior of the moduli space, far from the localization locus at Riemann surface degeneration. This has been studied  in heterotic string field theory \cite{Erbin:2019spp} and indeed shown to happen. This could also possibly reveal new localization channels associated with closed string degeneration. 

\item It would be also interesting to extend the localization mechanism to higher orders to understand possible underlying patterns.

\end{itemize}

In overall terms, including \cite{OC-I, OC-II}, we hope our works will trigger new exciting research on the physics of open and closed strings and on the non perturbative structure of string theory.
\section*{Acknowledgments}
We thank Ted Erler and Martin Schnabl for discussions and Ashoke Sen for correspondence.
We thank the organizers of ``Fundamental Aspects of String Theory'', Sao Paolo 1-12 June 2020,  and in particular Nathan Berkovits for giving us the opportunity to present some of our results prior to publication.
CM thanks CEICO and the Czech Academy of Science for hospitality during part of this work. JV also thanks INFN Turin for their hospitality during the initial stages of this work.  
The work of JV was supported by the NCCR SwissMAP that is funded by the Swiss National Science Foundation.
The work of CM is partially supported by the MIUR PRIN
Contract 2015MP2CX4 ``Non-perturbative Aspects Of Gauge Theories And Strings''.

\appendix

\section{Detailed calculations}
\label{app:proofs}

Here we give some detailed calculations and proofs of various results used in the main body of this paper.
\subsection{Proof of \eqref{eq:MassTermCorrA}}
\label{app:proofS11}

Here we will show in detail the equivalence of the leading-order mass-term correction in the $A_\infty$ and WZW-like open superstring field theories assuming the projector conditions $Q\psi=0=P_0e$. We will notice that our calculation is analogous to (but somewhat simpler than) the calculation of the coupling $\tilde{S}_{3,0}(\psi)$ presented in \cite{Mattiello:2019gxc} and \cite{Vosmera:2019mzw}. The projector condition $P_0e=0$ then turns out to play a role somewhat similar to that of the projector condition $P_0 M_2(\psi,\psi)=0$. 

Let us start with the expression \eqref{eq:S11} for $\tilde{S}_{1,1}(\psi)$. Going to the large Hilbert space (for the sake of computational convenience), substituting for $E_1$ in terms of $\mu_2$ and $e$ from \cite{OC-II} and using cyclicity of $M_2$, we can first rewrite this in the large Hilbert space as
	\begin{align}
	\tilde{S}_{1,1}(\psi) 
	&= \frac{1}{2}\omega_{\text{L}}(\xi_0\psi,\mu_2(\psi,e))+\frac{1}{2}\omega_{\text{L}}(\xi_0\psi,\mu_2(e,\psi))-\omega_{\text{L}}(\xi_0 M_2( \psi,\psi),h e)\,.\label{eq:step1}
\end{align}
Next we note that the third term of \eqref{eq:step1} can be rewritten as
\begin{align}
-\omega_{\text{L}}(\xi_0 M_2( \psi,\psi),h e) = \frac{1}{2}\omega_{\text{L}}( \xi_0 \psi,M_2(\psi,h e))+\frac{1}{2}\omega_{\text{L}}(\xi_0\psi,M_2(h e,\psi))\,.
\end{align}
Substituting the standard relation $M_2=[Q,\mu_2]$ into the third term of \eqref{eq:step1}, using cyclicity of $\mu_2$ and $Q\psi=Qe=0$, as well as the Hodge-Kodaira relation for the propagator $h$, as well as the projector condition $P_0e=0$,
we eventually obtain
\begin{subequations}
\begin{align}
	\tilde{S}_{1,1}(\psi)
	&= +\frac{1}{2}\omega_{\text{L}}(\xi_0\psi,\mu_2(\psi,e))+\frac{1}{2}\omega_{\text{L}}(\xi_0\psi,\mu_2(e,\psi))+\nonumber\\
	&\hspace{2cm}+
	\frac{1}{2}\omega_{\text{L}}( \xi_0 \psi,Q\mu_2(\psi,h e))+\frac{1}{2}\omega_{\text{L}}(\xi_0\psi,Q\mu_2(h e,\psi))+\nonumber\\
	&\hspace{2cm}-\frac{1}{2}\omega_{\text{L}}( \xi_0 \psi,\mu_2(\psi, e))-\frac{1}{2}\omega_{\text{L}}(\xi_0\psi,\mu_2( e,\psi))\\
	&= 	\frac{1}{2}\omega_{\text{L}}( \psi, X_0\mu_2(\psi,h e))+\frac{1}{2}\omega_{\text{L}}(\psi,X_0\mu_2(h e,\psi))
\end{align}
\end{subequations}
We now notice that it is possible to rewrite
\begin{subequations}
	\label{eq:rewrite}
\begin{align}
	\frac{1}{2}\omega_{\text{L}}( \psi, X_0\mu_2(\psi,h e)) &= -\frac{1}{4}\omega_{\text{L}}( \psi, X_0m_2(\xi_0\psi,h e))+\frac{1}{4}\omega_{\text{L}}( \xi_0\psi, X_0m_2(\psi, h e))\,,\\
		\frac{1}{2}\omega_{\text{L}}( \psi, X_0\mu_2(h e,\psi)) &= +\frac{1}{4}\omega_{\text{L}}( \xi_0\psi, X_0m_2(h e,\psi))-\frac{1}{4}\omega_{\text{L}}( \psi, X_0m_2( h e,\xi_0 \psi))\,,
\end{align}
\end{subequations}
because the string fields
\begin{subequations}
\begin{align}
	&\mu_2(\psi,he)+\frac{1}{2} m_2(\xi_0\psi,he)-\frac{1}{2} \xi_0 m_2(\psi,he)\,,\\
	&\mu_2(he,\psi)-\frac{1}{2} \xi_0 m_2(he,\psi)+\frac{1}{2} m_2( he,\xi_0 \psi)\,,
\end{align}
\end{subequations}
are clearly in the small Hilbert space. Substituting then $X_0 = Q\xi_0 +\xi_0 Q$ in the terms on the r.h.s.\ of \eqref{eq:rewrite} and moving the BRST operator $Q$ around (using again that $P_0 e=0$ as well as that $[m_2,e]=0$), we eventually obtain
\begin{subequations}
\label{eq:rewrite2}
\begin{align}
	\tilde{S}_{1,1}(\psi) &=-\frac{1}{4}\omega_{\text{L}}( \psi, \xi_0 Qm_2(\xi_0\psi,h e))+\frac{1}{4}\omega_{\text{L}}( \xi_0\psi, Q\xi_0 m_2(\psi, h e))+\nonumber\\
	&\hspace{6cm}+\frac{1}{4}\omega_{\text{L}}( \xi_0\psi, \xi_0 Qm_2(\psi, h e))+\nonumber\\
	&\hspace{1cm}+\frac{1}{4}\omega_{\text{L}}( \xi_0\psi, Q\xi_0 m_2(h e,\psi))
	+\frac{1}{4}\omega_{\text{L}}( \xi_0\psi, \xi_0 Qm_2(h e,\psi))+\nonumber\\
	&\hspace{6cm}
	-\frac{1}{4}\omega_{\text{L}}( \psi, \xi_0 Qm_2( h e,\xi_0 \psi))\\
	&=+\frac{1}{4}\omega_{\text{L}}( \psi, \xi_0 m_2(X_0\psi,h e))
	-\frac{1}{4}\omega_{\text{L}}( \psi, \xi_0 m_2(\xi_0\psi, e))+\nonumber\\
	&\hspace{2cm}	+\frac{1}{4}\omega_{\text{L}}( X_0\psi, \xi_0 m_2(\psi, h e))-\frac{1}{4}\omega_{\text{L}}( \xi_0\psi, \xi_0 m_2(\psi,  e))+\nonumber\\
	&\hspace{1cm}+\frac{1}{4}\omega_{\text{L}}( X_0\psi, \xi_0 m_2(h e,\psi))
	-\frac{1}{4}\omega_{\text{L}}( \xi_0\psi, \xi_0 m_2(e,\psi))
	+\nonumber\\
	&\hspace{2cm}
	+\frac{1}{4}\omega_{\text{L}}( \psi, \xi_0 m_2(  e,\xi_0 \psi))	+\frac{1}{4}\omega_{\text{L}}( \psi, \xi_0 m_2( h e,X_0 \psi))\\
		&=+\frac{1}{4}\omega_{\text{L}}( \psi, \xi_0 m_2(X_0\psi,h e))
	+\frac{1}{4}\omega_{\text{L}}( X_0\psi, \xi_0 m_2(\psi, h e))+\nonumber\\
	&\hspace{1cm}+\frac{1}{4}\omega_{\text{L}}( X_0\psi, \xi_0 m_2(h e,\psi))	+\frac{1}{4}\omega_{\text{L}}( \psi, \xi_0 m_2( h e,X_0 \psi))\,.
\end{align}
\end{subequations}
Using cyclicity of $m_2$, we can therefore write
\begin{align}
	\tilde{S}_{1,1}(\psi)	&=+\frac{1}{2}\omega_{\text{S}}( m_2(\psi,  X_0\psi),h e)
	+\frac{1}{2}\omega_{\text{S}}( m_2(X_0\psi,  \psi), h e)\,.
\end{align}

\subsection{Proof of \eqref{eq:tS02}}
\label{app:proofS02}

Paralleling the calculation in Appendix \eqref{app:proofS11}, we will now verify that the effective coupling $\tilde{S}_{0,2}(\psi)$ obtained from the $A_\infty$ theory can be rewritten in a way which is suitable for the comparison with the analogous prediction of the WZW-like theory, assuming that lower order couplings vanish by virtue of the conditions $P_0 e= P_0Q\psi=0$. Starting with the expression
\begin{align}
\tilde{S}_{0,2}(\psi) 
&= -\omega_\text{S}(\psi,E_1(he))+\omega_\text{S}(\psi,M_2(he,he))\,,
\end{align}
we can first rewrite this in the large Hilbert space as
\begin{align}
\tilde{S}_{0,2}(\psi) 
&= -\omega_\text{L}(\xi_0\psi,\mu_2(e,he))-\omega_\text{L}(\xi_0\psi,\mu_2(he,e))-\omega_\text{L}(\xi_0\psi,M_2(he,he))\,.
\end{align}
Substituting the standard relation $M_2=[Q,\mu_2]$ into the third term, using cyclicity of $\mu_2$ and $Qe=P_0 e=0$, as well as the Hodge-Kodaira relation for the propagator $h$, 
we can write
\begin{align}
-\omega_\text{L}(\xi_0\psi,M_2(he,he))&=-\omega_\text{L}(\xi_0\psi,Q\mu_2(he,he))+\nonumber\\
&\hspace{1cm}
+\omega_\text{L}(\xi_0\psi,\mu_2 ( e,h e))
+\omega_\text{L}(\xi_0\psi,\mu_2 (h e, e))\,.
\end{align}
so that the coupling can be rewritten as
\begin{subequations}
	\begin{align}
	\tilde{S}_{0,2}(\psi) 
	&= -\omega_\text{L}(\xi_0\psi,Q\mu_2(h e,h e))\\
	&= -\omega_\text{L}(\psi,X_0\mu_2(h e,h e))
	\end{align}
\end{subequations}
We now notice that it is possible to rewrite 
\begin{align}
-\omega_\text{L}(\psi,X_0\mu_2(h e,h e)) &= \frac{1}{2}\omega_\text{L}(\psi,X_0m_2(\xi_0 h e,h e))+\nonumber\\
&\hspace{3cm}+\frac{1}{2}\omega_\text{L}(\psi,X_0m_2(h e,\xi_0 h e)) \,,
\end{align}
because the string field
\begin{align}
\mu_2(h e,h e)+\frac{1}{2} m_2(\xi_0h e,h e)+\frac{1}{2} m_2(h e,\xi_0h e)
\end{align}
is clearly in the small Hilbert space. Substituting then $X_0 = Q\xi_0 +\xi_0 Q$ in both terms on the r.h.s.\ and moving the BRST operator $Q$ around (using again that $P_0 e=0$), we eventually obtain
\begin{subequations}
	\begin{align}
	\tilde{S}_{0,2}(\psi) 
	&= \frac{1}{2}\omega_\text{L}(\psi,\xi_0 Qm_2(\xi_0 h e,h e))+\nonumber\\
	&\hspace{3cm}+\frac{1}{2}\omega_\text{L}(\psi,\xi_0 Qm_2(h e,\xi_0 h e)) \\
	&= -\frac{1}{2}\omega_\text{L}(\xi_0\psi, m_2(Q\xi_0 h e,h e))+\frac{1}{2}\omega_\text{L}(\xi_0\psi, m_2(\xi_0 h e, e))+\nonumber\\
	&\hspace{1cm}-\frac{1}{2}\omega_\text{L}(\xi_0\psi, m_2( e,\xi_0 h e)) -\frac{1}{2}\omega_\text{L}(\xi_0\psi, m_2(h e,Q\xi_0 h e)) \,.
	\end{align}
\end{subequations}
Finally, using that $[{m}_2,{e}_0]=0$, the second and the third term cancel, so that we can write
\begin{subequations}
	\begin{align}
	\tilde{S}_{0,2}(\psi) 
	&= -\frac{1}{2}\omega_\text{L}(\xi_0\psi, m_2(Q\xi_0 h e,h e))\nonumber\\
	&\hspace{3cm}-\frac{1}{2}\omega_\text{L}(\xi_0\psi, m_2(h e,Q\xi_0 h e)) \\
	&= +\frac{1}{2}\big\langle\xi_0 \psi, [
	Q\xi_0 h e,h e]\big\rangle\,.
	\end{align}
\end{subequations}
This is then easily compared with the corresponding result in the WZW-like theory.

\subsection{Proof of \eqref{eq:S11loc}}
\label{app:proofloc}

Starting with the large Hilbert space form \eqref{eq:S11Berk} of the $\mathcal{O}(\mu)$ mass-term correction, we can first use the $\mathcal{N}=2$ $R$-charge conservation and $c$-ghost saturation to decompose
\begin{align}
	\tilde{{S}}_{1,1}= \tilde{{S}}_{1,1}^{\pm\pm}+\tilde{{S}}_{1,1}^{\pm\mp}\,,
\end{align}
where we define
\begingroup
\allowdisplaybreaks
\begin{subequations}
	\begin{align}
		\tilde{{S}}_{1,1}^{\pm\pm}&\equiv+\frac{1}{2}\varepsilon_{ij}\bigg\langle[\eta \varphi^+, Q\varphi^+ ],\xi_0
		\frac{b_0}{L_0}\bar{P}_0[\eta(\mathcal{U}^i)^-{Q}({\mathcal{U}}^j)^-]I\bigg\rangle+\nonumber\\
		&\hspace{2cm}+\frac{1}{2}\varepsilon_{ij}\bigg\langle[ \eta \varphi^+,Q\varphi^+ ]
		,\xi_0\frac{b_0}{L_0}\bar{P}_0[Q(\mathcal{U}^i)^-{\eta}({\mathcal{U}}^j)^-]I\bigg\rangle+\nonumber\\
		&\hspace{0.4cm}+\frac{1}{2}\varepsilon_{ij}\bigg\langle[\eta \varphi^-, Q\varphi^- ],\xi_0
		\frac{b_0}{L_0}\bar{P}_0[\eta(\mathcal{U}^i)^+{Q}({\mathcal{U}}^j)^+]I\bigg\rangle+\nonumber\\
		&\hspace{2cm}+\frac{1}{2}\varepsilon_{ij}\bigg\langle[ \eta\varphi^-,Q\varphi^- ]
		,\xi_0\frac{b_0}{L_0}\bar{P}_0[Q(\mathcal{U}^i)^+{\eta}({\mathcal{U}}^j)^+]I\bigg\rangle\,,\\
		\tilde{{S}}_{1,1}^{\pm\mp}&\equiv+\frac{1}{2}\varepsilon_{ij}\bigg\langle[\eta \varphi^+, Q\varphi^- ],\xi_0
		\frac{b_0}{L_0}\bar{P}_0[\eta(\mathcal{U}^i)^-{Q}({\mathcal{U}}^j)^+]I\bigg\rangle+\nonumber\\
		&\hspace{2cm}+\frac{1}{2}\varepsilon_{ij}\bigg\langle[ \eta \varphi^+,Q\varphi^- ]
		,\xi_0\frac{b_0}{L_0}\bar{P}_0[Q(\mathcal{U}^i)^+{\eta}({\mathcal{U}}^j)^-]I\bigg\rangle+\nonumber\\
		&\hspace{0.4cm}+\frac{1}{2}\varepsilon_{ij}\bigg\langle[\eta \varphi^-, Q\varphi^+ ],\xi_0
		\frac{b_0}{L_0}\bar{P}_0[\eta(\mathcal{U}^i)^+{Q}({\mathcal{U}}^j)^-]I\bigg\rangle+\nonumber\\
		&\hspace{2cm}+\frac{1}{2}\varepsilon_{ij}\bigg\langle[ \eta \varphi^-,Q\varphi^+ ]
		,\xi_0\frac{b_0}{L_0}\bar{P}_0[Q(\mathcal{U}^i)^-{\eta}({\mathcal{U}}^j)^+]I\bigg\rangle\,.
	\end{align}
\end{subequations}
\endgroup
First, note that we can rewrite
\begingroup
\allowdisplaybreaks
\begin{subequations}
	\begin{align}
		&\varepsilon_{ij}\bigg\langle[\eta \varphi^\pm, Q\varphi^\pm ],\xi_0
		\frac{b_0}{L_0}\bar{P}_0[\eta(\mathcal{U}^i)^\mp{Q}({\mathcal{U}}^j)^\mp]I\bigg\rangle=\nonumber\\
		&\hspace{2cm}=
		-\varepsilon_{ij}\bigg\langle[ \varphi^\pm, Q\varphi^\pm ],
		\frac{b_0}{L_0}\bar{P}_0 Q[\eta(\mathcal{U}^i)^\mp({\mathcal{U}}^j)^\mp]I\bigg\rangle\\
		&\hspace{2cm}=
		\varepsilon_{ij}\big\langle[ \varphi^\pm, Q\varphi^\pm ],
		{P}_0 [\eta(\mathcal{U}^i)^\mp({\mathcal{U}}^j)^\mp]I\big\rangle\nonumber\\[2mm]
		&\hspace{4cm}-\varepsilon_{ij}\big\langle[ \varphi^\pm, Q\varphi^\pm ],
		[\eta(\mathcal{U}^i)^\mp({\mathcal{U}}^j)^\mp]I\big\rangle\nonumber\\
		&\hspace{4cm}+\varepsilon_{ij}\bigg\langle[Q \varphi^\pm, Q\varphi^\pm ],
		\frac{b_0}{L_0}\bar{P}_0 [\eta(\mathcal{U}^i)^\mp({\mathcal{U}}^j)^\mp]I\bigg\rangle\label{eq:intstep2}\\
		&\hspace{2cm}=
		+\varepsilon_{ij}\big\langle[ \varphi^\pm, Q\varphi^\pm ],
		{P}_0 [\eta(\mathcal{U}^i)^\mp({\mathcal{U}}^j)^\mp]I\big\rangle\,,
	\end{align}
\end{subequations}
\endgroup
where we have first moved $\eta$ to cancel $\xi_0$, then moved one of the $Q$s through the propagator (using the Hodge-Kodaira decomposition in the process) and finally, in the last equality we have used $R$-charge conservation (together with the ghost number saturation) as well as the fact that the closed string state $\eta(\mathcal{U}^i)^\mp({\mathcal{U}}^j)^\mp$ is inserted at midpoint (which ensures vanishing of the contact term in \eqref{eq:intstep2}).
Similarly, we obtain
\begin{align}
	&\varepsilon_{ij}\bigg\langle[ \eta \varphi^\pm,Q\varphi^\pm ]
	,\xi_0\frac{b_0}{L_0}\bar{P}_0[Q(\mathcal{U}^i)^\mp{\eta}({\mathcal{U}}^j)^\mp ]I\bigg\rangle=\nonumber\\
	&\hspace{4cm}=-\varepsilon_{ij}\big\langle[  \varphi^\pm,Q\varphi^\pm ]
	,{P}_0[(\mathcal{U}^i)^\mp{\eta}({\mathcal{U}}^j)^\mp ]I\big\rangle\,.
\end{align}
Completely analogous manipulations can also be applied to rewrite $\tilde{{S}}_{11}^{\pm\mp}$. Eventually, we obtain
\begingroup
\allowdisplaybreaks
\begin{subequations}
	\begin{align}
		&\varepsilon_{ij}\bigg\langle[\eta \varphi^\pm, Q\varphi^\mp ],\xi_0
		\frac{b_0}{L_0}\bar{P}_0[\eta(\mathcal{U}^i)^\mp {Q}({\mathcal{U}}^j)^\pm]I\bigg\rangle=\nonumber\\
		&\hspace{6cm}=+\varepsilon_{ij}\big\langle[\eta \varphi^\pm, Q\varphi^\mp ],
		{P}_0 [(\mathcal{U}^i)^\mp({\mathcal{U}}^j)^\pm]I\big\rangle\,,\\
		&\varepsilon_{ij}\bigg\langle[ \eta\varphi^\pm,Q\varphi^\mp ]
		,\xi_0\frac{b_0}{L_0}\bar{P}_0[Q(\mathcal{U}^i)^\pm{\eta}({\mathcal{U}}^j)^\mp ]I\bigg\rangle=\nonumber\\
		&\hspace{6cm}=-\varepsilon_{ij}\big\langle[ \eta \varphi^\pm,Q\varphi^\mp ]
		,{P}_0[(\mathcal{U}^i)^\pm ({\mathcal{U}}^j)^\mp ]I\big\rangle\,.
	\end{align}
\end{subequations}
\endgroup
Altogether we have therefore established that $\tilde{{S}}_{1,1}$ can be rewritten in a purely localized form
\begin{align}
	\tilde{{S}}_{1,1}
	&=-\frac{1}{2}\varepsilon_{ij}\big\langle\big([ \eta \varphi^+,Q\varphi^- ]\!-\![ \eta \varphi^-,Q\varphi^+ ]\big)
	,\!{P}_0\big[(\mathcal{U}^i)^+ ({\mathcal{U}}^j)^- \!-\!(\mathcal{U}^i)^- ({\mathcal{U}}^j)^+\big]I\big\rangle\nonumber\\
	&\hspace{2.5cm}-\frac{1}{2}\varepsilon_{ij}\big\langle[  \varphi^+,Q\varphi^+ ]
	,{P}_0\big[(\mathcal{U}^i)^-{\eta}({\mathcal{U}}^j)^- - {\eta}(\mathcal{U}^i)^-({\mathcal{U}}^j)^-\big]I\big\rangle\nonumber\\
	&\hspace{2.5cm}-\frac{1}{2}\varepsilon_{ij}\big\langle[  \varphi^-,Q\varphi^- ]
	,{P}_0\big[(\mathcal{U}^i)^+{\eta}({\mathcal{U}}^j)^+ - {\eta}(\mathcal{U}^i)^+({\mathcal{U}}^j)^+\big]I\big\rangle\,.
\end{align}

\endgroup

\begin{thebibliography}{99}

%

 \bibitem{OC-I}
C.~Maccaferri and J.~Vo\v{s}mera,
``Closed string deformations in open string field theory I: bosonic string,''
[arXiv:2103.04919 [hep-th]].
 
 \bibitem{OC-II}
C.~Maccaferri and J.~Vo\v{s}mera,
``Closed string deformations in open string field theory II: superstring,''
[arXiv:2103.04920 [hep-th]].
 
  \bibitem{Berkovits}
  N.~Berkovits,
  ``SuperPoincare invariant superstring field theory,''
  Nucl.\ Phys.\ B {\bf 450} (1995) 90
   Erratum: [Nucl.\ Phys.\ B {\bf 459} (1996) 439]
  doi:10.1016/0550-3213(95)00620-6, 10.1016/0550-3213(95)00259-U
  [hep-th/9503099].

\bibitem{Berkovits:2004xh}
  N.~Berkovits, Y.~Okawa and B.~Zwiebach,
  ``WZW-like action for heterotic string field theory,''
  JHEP {\bf 0411} (2004) 038
  doi:10.1088/1126-6708/2004/11/038
  [hep-th/0409018].

 
\bibitem{Erler:2013xta}
T.~Erler, S.~Konopka and I.~Sachs,
``Resolving Witten`s superstring field theory,''
JHEP \textbf{04} (2014), 150
doi:10.1007/JHEP04(2014)150
[arXiv:1312.2948 [hep-th]].

 \bibitem{Maccaferri:2019ogq}
  C.~Maccaferri and A.~Merlano,
  ``Localization of effective actions in open superstring field theory: small Hilbert space,''
  JHEP {\bf 1906} (2019) 101
  doi:10.1007/JHEP06(2019)101
  [arXiv:1905.04958 [hep-th]].
  
\bibitem{Billo:2008sp}
M.~Billo, L.~Ferro, M.~Frau, F.~Fucito, A.~Lerda and J.~F.~Morales,
``Flux interactions on D-branes and instantons,''
JHEP \textbf{10} (2008), 112
doi:10.1088/1126-6708/2008/10/112
[arXiv:0807.1666 [hep-th]].


\bibitem{Larocca:2017pbo}
P.~V.~Larocca and C.~Maccaferri,
``BCFT and OSFT moduli: an exact perturbative comparison,''
Eur. Phys. J. C \textbf{77} (2017) no.11, 806
doi:10.1140/epjc/s10052-017-5379-3
[arXiv:1702.06489 [hep-th]].



  \bibitem{Sen:2019jpm}
  A.~Sen,
  ``String Field Theory as World-sheet UV Regulator,''
  JHEP {\bf 1910} (2019) 119
  doi:10.1007/JHEP10(2019)119
  [arXiv:1902.00263 [hep-th]].

\bibitem{Sen:2020cef}
A.~Sen,
``D-instanton Perturbation Theory,''
[arXiv:2002.04043 [hep-th]].



   \bibitem{Sen-restoration}
  A.~Sen,
  ``Supersymmetry Restoration in Superstring Perturbation Theory,''
  JHEP {\bf 1512} (2015) 075
  doi:10.1007/JHEP12(2015)075
  [arXiv:1508.02481 [hep-th]].

\bibitem{Erbin:2019spp}
H.~Erbin, C.~Maccaferri and J.~Vošmera,
``Localization of effective actions in Heterotic String Field Theory,''
JHEP \textbf{02} (2020), 059
doi:10.1007/JHEP02(2020)059
[arXiv:1912.05463 [hep-th]].
 
  \bibitem{Maccaferri:2018vwo}
  C.~Maccaferri and A.~Merlano,
  ``Localization of effective actions in open superstring field theory,''
  JHEP {\bf 1803} (2018) 112
  doi:10.1007/JHEP03(2018)112
  [arXiv:1801.07607 [hep-th]].
  
  \cite{Erler:2015rra}
\bibitem{Erler:2015rra}
T.~Erler, Y.~Okawa and T.~Takezaki,
``$A_\infty$ structure from the Berkovits formulation of open superstring field theory,''
[arXiv:1505.01659 [hep-th]].

\bibitem{Erler:2015uoa}
T.~Erler,
``Relating Berkovits and $A_\infty$ superstring field theories; large Hilbert space perspective,''
JHEP \textbf{02} (2016), 121
doi:10.1007/JHEP02(2016)121
[arXiv:1510.00364 [hep-th]].

\bibitem{Erler:2015uba}
T.~Erler,
``Relating Berkovits and A$_\infty$ superstring field theories; small Hilbert space perspective,''
JHEP \textbf{10} (2015), 157
doi:10.1007/JHEP10(2015)157
[arXiv:1505.02069 [hep-th]].

  
  
  














 
 
%




  \bibitem{Vosmera:2019mzw}
  J.~Vo\v{s}mera,
  ``Generalized ADHM equations from marginal deformations in open superstring field theory,''
  arXiv:1910.00538 [hep-th].
  
  
  
\bibitem{Billo:2021xzh}
M.~Billo, M.~Frau, F.~Fucito, L.~Gallot, A.~Lerda and J.~F.~Morales,
``On the D(-1)/D7-brane systems,''
[arXiv:2101.01732 [hep-th]].

  
  
  

 
%


%

 


   
  

  
%
%
  
\bibitem{Billo:2005fg}
M.~Billo, M.~Frau, S.~Sciuto, G.~Vallone and A.~Lerda,
``Non-commutative (D)-instantons,''
JHEP \textbf{05} (2006), 069
doi:10.1088/1126-6708/2006/05/069
[arXiv:hep-th/0511036 [hep-th]].




\bibitem{David:2000um}
J.~R.~David,
``Tachyon condensation in the D0 / D4 system,''
JHEP \textbf{10} (2000), 004
doi:10.1088/1126-6708/2000/10/004
[arXiv:hep-th/0007235 [hep-th]].

\bibitem{Obers:1998fb}
N.~A.~Obers and B.~Pioline,
``U duality and M theory,''
Phys. Rept. \textbf{318} (1999), 113-225
doi:10.1016/S0370-1573(99)00004-6
[arXiv:hep-th/9809039 [hep-th]].




\bibitem{Mattiello:2019gxc}
  L.~Mattiello and I.~Sachs,
  ``On Finite-Size D-Branes in Superstring Theory,''
  JHEP {\bf 1911} (2019) 118
  doi:10.1007/JHEP11(2019)118
  [arXiv:1902.10955 [hep-th]].


\bibitem{David:2002wn}
J.~R.~David, G.~Mandal and S.~R.~Wadia,
``Microscopic formulation of black holes in string theory,''
Phys. Rept. \textbf{369} (2002), 549-686
doi:10.1016/S0370-1573(02)00271-5
[arXiv:hep-th/0203048 [hep-th]].

\bibitem{Seiberg:1999vs}
N.~Seiberg and E.~Witten,
``String theory and noncommutative geometry,''
JHEP \textbf{09} (1999), 032
doi:10.1088/1126-6708/1999/09/032
[arXiv:hep-th/9908142 [hep-th]].

\bibitem{Douglas:1996sw}
M.~R.~Douglas and G.~W.~Moore,
``D-branes, quivers, and ALE instantons,''
[arXiv:hep-th/9603167 [hep-th]].

\bibitem{Nekrasov:1998ss}
N.~Nekrasov and A.~S.~Schwarz,
``Instantons on noncommutative R**4 and (2,0) superconformal six-dimensional theory,''
Commun. Math. Phys. \textbf{198} (1998), 689-703
doi:10.1007/s002200050490
[arXiv:hep-th/9802068 [hep-th]].


\bibitem{Schnabl:2019oom}
M.~Schnabl and J.~Vo\v{s}mera,
``Gepner-like boundary states on $T^4$,''
[arXiv:1903.00487 [hep-th]].










%

\bibitem{Erbin:2020eyc}
H.~Erbin, C.~Maccaferri, M.~Schnabl and J.~Vo\v{s}mera,
``Classical algebraic structures in string theory effective actions,''
JHEP \textbf{11} (2020), 123
doi:10.1007/JHEP11(2020)123
[arXiv:2006.16270 [hep-th]].

\bibitem{Koyama:2020qfb}
D.~Koyama, Y.~Okawa and N.~Suzuki,
``Gauge-invariant operators of open bosonic string field theory in the low-energy limit,''
[arXiv:2006.16710 [hep-th]].


  

  
 
%

  

%
%
 
 \bibitem{Moosavian:2019ydz}
  S.~Faroogh Moosavian, A.~Sen and M.~Verma,
  ``Superstring Field Theory with Open and Closed Strings,''
  arXiv:1907.10632 [hep-th].

%
%

%

%


%
%

%

%
%
%
%
%
%










 \end{thebibliography}
\end{document}